\documentclass[journal]{IEEEtran}

\pdfoutput=1

\usepackage{amsmath,amsfonts}
\usepackage{algorithmic}
\usepackage{algorithm}
\usepackage{array}
\usepackage{textcomp}
\usepackage{stfloats}
\usepackage{verbatim}
\usepackage{graphicx}
\usepackage{cite}
\usepackage{xurl} 
\usepackage{enumitem}
\usepackage{amsmath}
\usepackage{threeparttable}
\usepackage{array}
\usepackage{makecell}
\usepackage{multirow}
\usepackage{booktabs}
\usepackage[normalem]{ulem}
\usepackage{graphicx}
\usepackage{tabularx}
\usepackage{amsmath}

\usepackage[most]{tcolorbox}
\usepackage{listings}
\usepackage{xcolor}

\lstdefinestyle{pythonstyle}{
    basicstyle=\ttfamily\scriptsize,
    breaklines=true,
    columns=fullflexible,
    frame=none,
    keepspaces=true,
    showstringspaces=false,
    aboveskip=2pt,
    belowskip=2pt
}

\newtcolorbox{dialoguebox}[1]{
    colback=gray!4,
    colframe=gray!45,
    title=#1,
    fonttitle=\bfseries\scriptsize,
    coltitle=black,
    boxrule=0.4pt,
    arc=1mm,
    left=1mm,
    right=1mm,
    top=1mm,
    bottom=1mm,
    before skip=4pt,
    after skip=4pt
}

\ifCLASSOPTIONcompsoc
    \usepackage[caption=false,font=normalsize,labelfont=sf,textfont=sf]{subfig}
\else
    \usepackage[caption=false,font=footnotesize]{subfig}
\fi

\usepackage[most]{tcolorbox}
\tcbset{
  my box/.style={
    enhanced,
    colframe=#1!80,
    colback=#1!10,
    attach boxed title to top left={xshift=0.2cm, yshift=-0.2cm},
    boxed title style={
      colback=#1!80,
      outer arc=0pt,
      arc=0pt,
      top=0pt,
      bottom=0pt,
    },
  },
}

\AtBeginDocument{%
  }

\begin{document}

\title{Characterizing Readability Issue Patterns and the Role of Prompt Design in LLM-Generated Code}

\author{Hengzhi Ye, Fengyuan Ran, Weiwei Xu, Minghui Zhou

\thanks{Hengzhi Ye, Weiwei Xu and Minghui Zhou are with the School of Computer Science, Peking University,
Beijing 100871, China (e-mail: hzye@stu.pku.edu.cn, xuww@stu.pku.edu.cn, zhmh@pku.edu.cn). Fengyuan Ran is with Wuhan University, Wuhan 430072, China (e-mail: rfy\_reflow@whu.edu.cn).}

}

\maketitle

\begin{abstract}
Large Language Models (LLMs) are increasingly changing how code is produced, but generated code still requires human review and validation before it can be adapted or integrated into real-world projects. This makes the readability of LLM-generated code a critical concern.
Existing studies have mainly focused on functional correctness and task completion of LLM generated code, leaving open questions about whether it is readable, how its readability fails, and to what extent prompt design can improve it.
We therefore investigate the readability of LLM-generated code by examining how it compares with human-written code, what distinct issue patterns it exhibits, and how prompt design is associated with these readability outcomes.
We first construct a readability assessment model that integrates textual, structural, program, and visual features. Using this model, we compare human-written code with code generated by representative frontier LLMs across 2,735 scenarios derived from World of Code (WoC) and LeetCode. We further characterize readability issue patterns using thematic analysis and examine prompt design associations through controlled prompt-variant experiments.
Our results show that current LLMs produce code that is comparable to human-written code in overall readability. However, this aggregate similarity masks distinct issue patterns, including excessive complexity, redundant comments, and unknown API usage. Prompt analysis further shows that \textit{function signature}, \textit{constraints}, and \textit{style description} are most strongly associated with code readability, although the overall role of prompt design remains bounded.
These findings reveal latent readability debt in AI-assisted programming, identify prompt design as a lightweight starting point for improving generated-code readability, and motivate automated support for detecting and mitigating readability issues in future development workflows.
\end{abstract}


\section{Introduction}\label{sec:intro}

Code readability has long been critical to software engineering, as developers have to first understand code before they can review, modify, reuse, or extend it~\cite{elshoff1982improving}. Readable code reduces comprehension effort, lowers maintenance burden, and helps prevent technical debt~\cite{boswell2011art,digkas2020can}.

The widespread use of Large Language Models (LLMs) has changed how code is produced. However, LLM-generated code still needs human oversight, as it may contain hallucinations, outdated knowledge, or subtle vulnerabilities~\cite{liu2023your,zhong2024can}. Developers are therefore required to judge whether generated code can be trusted to be adapted or integrated into development workflows. When such code is hard to read, the cost of review may offset the productivity gains of AI assistance~\cite{zi2025would}.

Existing research has primarily evaluated LLM-generated code through functional correctness and task completion~\cite{coignion2024performance, chen2021evaluating}. Although essential, these criteria say little about how readable generated code really is once it enters human review and maintenance workflows. A large-scale comparison with human-written code is therefore important to establish a practical baseline. However, overall readability scores alone cannot reveal how readability fails, as LLM-generated code may be difficult to read for reasons different from human-written code. Furthermore, although prompt design is the primary way users steer LLM outputs, little is known about which prompt dimensions are associated with more readable generated code. 
Understanding these questions is important for improving the reliability and maintainability of AI-assisted software development.

Thus, this paper investigates the readability of LLM-generated code through three research questions:

\begin{itemize}[leftmargin=1.5em, labelsep=0.5em]
    \item \textbf{RQ1}: How does the overall readability of LLM-generated code compare with human-written code?
    \item \textbf{RQ2}: What readability issue patterns distinguish LLM-generated code from human-written code?
    \item \textbf{RQ3}: What role does prompt design play in shaping the readability of LLM-generated code?
\end{itemize}

We started by constructing a readability assessment model that integrates textual, structural, program, and visual features into a unified framework.
Using this model, we compare human-written code with code generated by representative frontier LLMs available at the time of our study across 2,735 scenarios extracted from World of Code (WoC)~\cite{ma2019world} and LeetCode~\cite{leetcode}. 
We then examined the readability issues LLM-generated code exhibits and studied whether their patterns differ from those in human-written code through a hybrid deductive–inductive thematic analysis.
Finally, we analyzed the relationship between prompt dimensions and generated-code readability across 5,248 samples produced from controlled prompt variants.

Our results show that LLM-generated code is broadly comparable to human-written code in overall readability. However, a critical divergence exists behind superficial similarity: LLM-generated code exhibits distinct readability issue patterns, including unnecessary complex structures, redundant comments, unknown API usage, etc. Further prompt analysis shows that \textit{function signature}, \textit{constraints}, and \textit{style description} are most associated with the readability of generated code, while the overall role of prompt design is bounded.
These findings reveal how LLM-generated code differs from human-written code in readability, expose latent readability debt in AI-assisted programming, and identify prompt design as a lightweight starting point for readability improvement, offering insights for managing and improving LLM-generated code.

In summary, this paper makes the following contributions:

\begin{itemize}[leftmargin=1.5em, labelsep=0.5em]
    \item An \textbf{empirical study} of LLM-generated code readability, showing that LLM-generated code is broadly comparable to human-written code in aggregate readability but differs in concrete readability issue patterns;
    \item An empirical \textbf{analysis of prompt design}, identifying the prompt dimensions associated with more readable code while showing that their role remains bounded;
    \item A \textbf{scalable benchmark} for comparing LLM-generated and human-written code, consisting of 2,735 task prompts and corresponding human implementations from real-world software practices and programming competitions;  
    \item A comprehensive \textbf{readability assessment model} that integrates textual, structural, program, and visual features to quantify code readability at scale.
\end{itemize}

In Section \ref{sec:bg}, we review related work. We illustrate the overall methodology in Section~\ref{sec:methodology} and explain our readability assessment model in Section~\ref{sec:readability}. Sections~\ref{sec:rq1} to \ref{sec:rq3} detail the results for each research question. We discuss implications and limitations in Section~\ref{sec:discussion} and conclude in Section \ref{sec:conclusion}.

\section{Related Work}\label{sec:bg}

\subsection{LLM Code Generation}\label{s:related-llm}

The application of LLMs to code generation substantially enhanced automated programming capabilities. Not only have general LLMs like ChatGPT been widely deployed for code generation~\cite{jin2024llms}, but fine-tuned models such as Codex~\cite{chen2021evaluating} and StarCoder~\cite{li2023starcoder} have also emerged to achieve competitive performance across various programming languages. Integrated code agents like Copilot~\cite{copilot} and Cursor~\cite{cursor} further enhance developer productivity. Surveys indicate that Copilot is behind an average of 46\% of developers' code across all programming languages~\cite{copilotreport}, highlighting the profound impact of LLM-generated code on software development. 

Researchers have focused on examining whether LLM-generated code is functionally correct, secure, and aligned with developer expectations. For example, HumanEval~\cite{chen2021evaluating} utilizes reference solutions and unit tests to quantify the functional correctness of LLM-generated code. Additional evaluation criteria have also examined LLM-generated code in terms of vulnerability risks, privacy protection, and copyright compliance~\cite{yao2024survey, xu2025licoeval}.
However, little attention is devoted to the readability of the generated code despite its importance to long-term software lifecycle. 

\subsection{Code Readability}~\label{s:related-readability}

Since the emergence of programming, code readability has become an implicit code quality criterion. 

Buse and Weimer~\cite{buse2008metric} developed the first readability model by mapping 25 structural features, such as indentation and line length, to human-annotated code. Their findings established a positive correlation between readability scores and overall code quality. Subsequent studies refined structural metrics, with Posnett et al.~\cite{posnett2011simpler} achieving higher accuracy through a simpler three-feature model and Johnson et al.~\cite{johnson2019empirical} exploring how nesting and looping constructs influenced readability. A significant advancement occurred as Scalabrino et al.~\cite{scalabrino2016improving, scalabrino2018comprehensive} integrated textual and semantic features with structural ones to capture the dual nature of code as both logic and language.

More recently, researchers have examined code readability within the LLM context. Takerngsaksiri et al.~\cite{takerngsaksiri2025code} explored practitioners’ perspectives on code readability in the age of LLMs through a case study, while Pan et al.~\cite{pan2025hidden} investigated the influence that readability has on the LLM budget when LLMs process code snippets. Furthermore, Hu et al.~\cite{hu2024effectively} investigated LLMs' performance when interpreting poor-readability code, finding that even LLMs struggle with low-readability code, which mirrors human developers' difficulties that lead to technical debt accumulation. These findings further emphasize the detrimental impact of low-readability code in software engineering.

\subsection{Prompt Engineering} 

As an efficient alternative to costly tuning and retraining, prompt engineering has emerged as a primary paradigm for optimizing LLM performance across diverse tasks~\cite{marvin2023prompt, white2023prompt}. Since prompt design directly shapes model outputs, it likely influences the readability of the generated code.

Leveraging their extensive pre-training, LLMs function as capable few-shot learners through zero-shot instructions or in-context examples~\cite{wei2021finetuned, perez2021true}. The adoption of prompt engineering across various domains has established task-specific best practices. Researchers have defined design principles for specialized contexts such as academic writing and business management~\cite{giray2023prompt, busch2023just}, while practical implementations like ``cursor rules''~\cite{cursorrule} utilize pre-set contexts to optimize code generation. These frameworks improve LLMs' output by explicitly shaping the input prompt~\cite{cursorrule, giray2023prompt, liu2023pre}, including persona settings, stylistic guidelines, function signature, and technical contracts. Beyond structural optimization, prompt engineering also incorporates reasoning-oriented methods, as Chain-of-Thought (CoT) prompting~\cite{wei2022chain} enhances performance by simulating step-by-step problem decomposition.

In summary, current research on LLM code generation primarily emphasizes functional capacity, and prompt engineering literature has prioritized strategies for enhancing functional performance, leaving the crucial non-functional dimension of readability unexplored. Evaluating the readability of LLM-generated code at scale, identifying specific issue patterns, and evaluating the impact of prompt design thus remain critical yet unaddressed knowledge gaps.

\section{Methodology} \label{sec:methodology}

\begin{figure*}[t]
    \centering
    \includegraphics[width=0.95\linewidth]{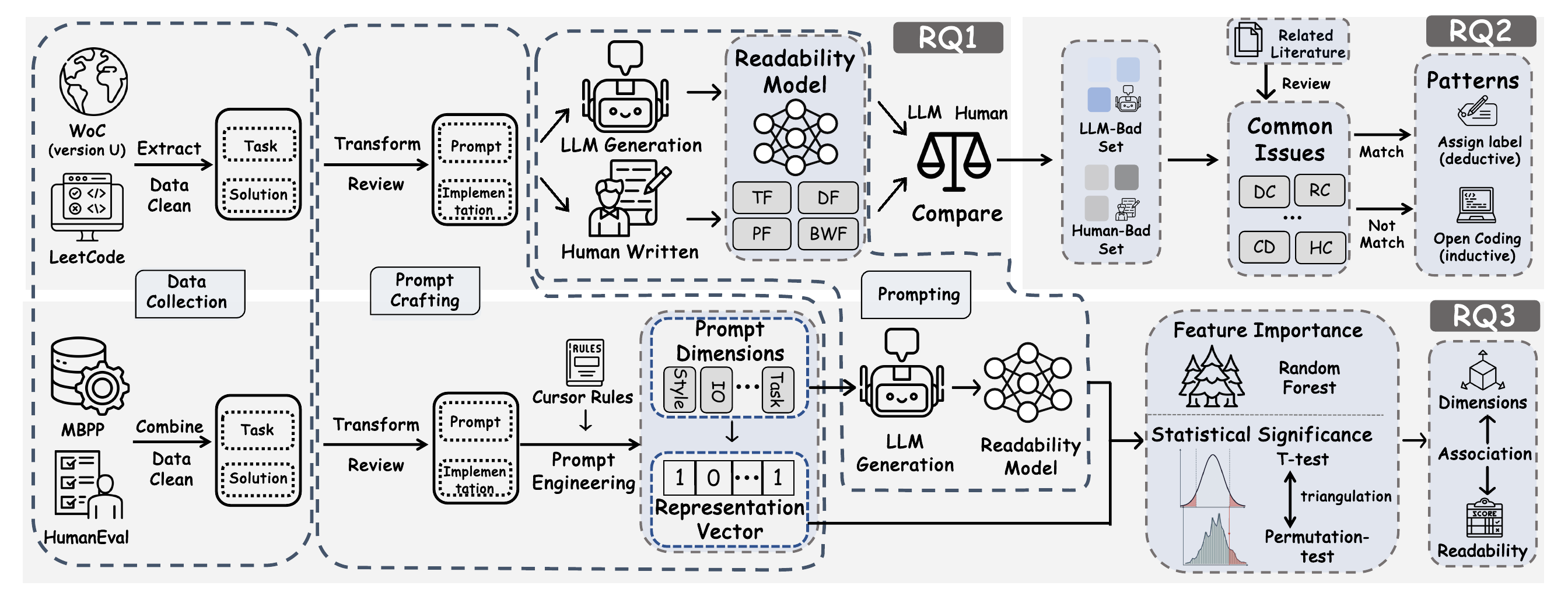}
    \vspace{-1em}
    \caption{Overview of the methodology}
    \vspace{-.5em}
    \label{fig:methodology}
\end{figure*}

As shown in Figure~\ref{fig:methodology}, we employed a mixed-method approach to address our research questions. 
We started by collecting and processing source data to obtain structured prompt and human implementation.
Then we prompted LLMs to generate code responses, comparing the readability of LLM-generated code with human-written code.
After that, we identified readability issue patterns through hybrid deductive–inductive thematic analysis, and investigated the impact of different prompt dimensions on generated code readability using statistical methods.

\subsection{Data Collection}

To make LLM application scenarios more representable, we selected both real software engineering development contexts and competitive programming problems as task backgrounds. Specifically, we used World of Code (WoC)~\cite{ma2019world} and LeetCode~\cite{leetcode} as data sources. Given that ChatGPT was released in November 2022~\cite{chatgpt}, before which LLMs were neither widely adopted nor integrated into programming workflows, we selected data snapshots prior to 2022 so that the collected code is human-written. 
Note that, although the pre-2022 data may overlap with the LLM's pre-training corpus, this does not compromise the validity of our conclusions. LLMs acquire code generation capabilities through next-token prediction conditioned on statistical patterns learned from large-scale training corpora, and their outputs inherently reflect these learned distributions over human-written code. Whether or not a problem was encountered during pre-training, the generated code draws from the same underlying distribution, which is what practitioners obtain when using LLMs in practice. Unlike studies that assess functional correctness, where overlap with training data may artificially inflate pass rates on known benchmarks, our evaluation targets stylistic and structural properties of the generated code, which remain meaningful regardless of whether the solution logic has been previously encountered.

WoC is a comprehensive infrastructure for mining version control data across open-source software ecosystem, which aggregates Git objects, including commits, trees and blobs, on platforms like GitHub, Bitbucket, and GitLab~\cite{ma2019world, ma2021world}. It provides mappings from project identification to project code, blobs, version control information, and other data for each project entity, facilitating easy querying and management. 

We selected version U of WoC (released in October 2021) and used the \texttt{c2fbb} database, which maps each commit to its corresponding commit file name, new blob (post-commit), and old blob (pre-commit), filtering for Python project files\footnote{To maintain a manageable research scope, in this study, we focused specifically on Python code readability.}. 
To mitigate the risk of including non-representative code (e.g., toy projects or experimental repositories), we applied the following selection criteria: (1) we retained only projects with at least 10 commits, ensuring a minimum level of development activity and maturity; (2) we excluded projects whose repository names or descriptions contained keywords indicative of academic exercises or tutorials (e.g., \textit{``homework,'' ``assignment,'' ``tutorial,'' ``demo,'' ``test,''} etc.); (3) following Buse and Weimer's definition of readability~\cite{buse2008metric}, we utilized function-level snippets and retained only functions between 5 and 100 lines of code (excluding blank lines and comments). This granularity ensures sufficient context without spanning multiple methods, while filtering out trivially short snippets and excessively long functions that could introduce confounding complexity.

We therefore extracted and segmented Python source files into individual function-level entities for analysis. We further divided each snippet into a function header and body: the header, containing the function signature and docstring description, would serve as the prompt to stimulate LLM code generation, while the body, representing the concrete implementation, would serve as human-written baseline. Additionally, we excluded functions whose docstrings were absent or shorter than 10 words, as an insufficiently descriptive header would constitute an ambiguous prompt that cannot reliably guide LLM generation.
After filtering and selection, we finally obtained 1,000 prompt-baseline pairs.

LeetCode~\cite{leetcode} is an online platform that provides programming problems spanning various difficulty levels and topics. It offers a standardized problem set with clear problem descriptions in natural languages, and worldwide users provide a diverse array of human-written code in multiple programming languages~\cite{coignion2024performance}.

We crawled all accessible problems from LeetCode, including their problem descriptions and code solutions, and filtered for solutions implemented in Python and released before 2022. We then applied the following selection criteria: (1) for each problem, when multiple solutions existed, we selected the most popular one with the highest number of upvotes, as this metric reflects peer-validated quality and idiomatic style; (2) consistent with the WoC dataset, we retained only solutions whose core function body ranged between 5 and 100 lines of code. Subsequently, we extracted the core function implementation from each code solution to serve as the human-written baseline, with the corresponding problem description serving as the prompt. 

Ultimately, we obtained 1,735 pairs from LeetCode. Combining with WoC-sourced data, we got 2,735 effective samples in total.

\subsection{Prompt Crafting}
To support our multi-stage investigation, we crafted two different prompt sets tailored to different research objectives. The \textbf{first set A} should provides a diverse array of real-world scenarios to evaluate the general readability of LLM-generated code and identify specific defect patterns \textbf{(RQ1 and RQ2)}, while the \textbf{second set B} should facilitate a controlled experimental environment for \textbf{RQ3}. This strategic separation allows us to maintain ecological validity for problem identification while ensuring the empirical rigor required for variable-controlled analysis of prompt dimensions.

\subsubsection{Set A: Real-World Scenario Prompts (RQ1 \& RQ2)}

This set establishes a baseline for general readability and defect identification. Specifically, we used the 2,735 samples previously collected from real software engineering practice as the foundational data, which includes 1,735 natural language problem descriptions from LeetCode and 1,000 function-level docstrings from WoC. 

The LeetCode problem descriptions are already self-contained natural language task specifications and were used directly as prompts without modification, while the WoC docstrings are inline reference annotations for existing implementations (e.g., parameter type lists, return value notes) rather than as task-oriented instructions to prompt code generation. Directly using such docstrings as prompts would produce inputs unrepresentative of how developers interact with LLMs in practice, where users formulate requests in instructional natural language~\cite{chen2026code}. To address this, two authors independently transformed the WoC docstrings into natural language task descriptions following three constraints: 

\begin{itemize}
    \item \textbf{Semantic preservation:} the functional requirement expressed must be strictly equivalent to the original docstring with no addition or omission of logic;
    \item \textbf{Minimal stylistic intervention:} the transformation adopts plain instructional phrasing without embellishing with stylistic or readability-related descriptions;
    \item \textbf{Being self-contained:} the resulting prompt must be interpretable without access to the surrounding codebase. 
\end{itemize}

This process involved independent transformation followed by a cross-validation phase in which each author reviewed the other's outputs against the three constraints. Any discrepancies were resolved through iterative discussion to a consensus on all final expressions.
We finally got a comprehensive set of 5,086 prompts.

\subsubsection{Set B: Controlled Experimental Prompts (RQ3)}

Analyzing how specific prompt dimensions affect readability requires a rigorous control of variables. The complexity of real software practice in the first set makes it difficult to decouple specific descriptive elements into isolated variables. Thus, we adopted a from-scratch construction strategy, prioritizing simplicity and modularity to ensure experimental control.

We sourced problems from MBPP~\cite{austin2021program} and HumanEval~\cite{chen2021evaluating} as they are established benchmarks with concise descriptions that are easily decoupled into base components. To maintain a balanced sample size, we combined 164 HumanEval samples with a randomly sampled subset of 164 MBPP problems. Similarly, natural language task descriptions originally provided by MBPP  were used as prompts without modification, while for HumanEval, whose prompts are formatted as embedded Python docstrings, we converted them into standalone natural language instructions following the same three constraints. 
This process was also followed by author cross-validation to ensure rigor and accuracy.

Referring to prior work~\cite{giray2023prompt,white2023prompt,zhang2025prompt}, we limited the dimensions of prompt crafting to 8 types, as shown in Table~\ref{tab:prompt_dimension}. 
\textit{Minimum task description} served as the baseline, consisting of the most basic problem description without any additional information. \textit{Function signature}, \textit{IO contract}, \textit{few-shot instruction}, and \textit{task category} were extracted from the code implementation and test cases, while \textit{style description}, \textit{persona setting}, and \textit{constraint} were derived from best practices in cursor rules~\cite{cursorrule}. 
Starting with each problem's baseline prompt, we constructed prompts that only added one dimension, omitted only one dimension, or included all dimensions and recorded dimensions that current prompt contained with a seven-element vector:

\[
V_{\text{prompt}} = (\mathbb{I}\{d\})_{d \in \mathcal{D}},\]

where $\mathcal{D} = \{$\text{style}, \text{signature}, \text{IO}, \text{few-shot},
\text{category}, \text{persona}, \text{constraint}$\}$, and $\mathbb{I}\{d\}=1$ if dimension $d$ is included in the prompt, otherwise $0$.

Finally, each baseline corresponded to 16 different prompts and vectors representing their crafting elements, ultimately yielding 5,248 operational (\textit{prompt, vector}) pairs in set B.

\subsection{Prompting and Data Processing}\label{subsec:prompt}

For the prompt set A (RQ1 \& RQ2), we input the 2,735 prompts into a set of prevalent LLMs, including \texttt{GPT-5}~\cite{singh2025openai}, \texttt{Claude-4.6}~\cite{claude}, \texttt{DeepSeek-v4}~\cite{deepseek}, \texttt{Gemini 3}~\cite{gemini} and \texttt{Qwen 3.5}~\cite{qwen}, to obtain their responses, with \textit{temperature} set to 0 and \textit{max\_tokens} set to 4,096.
We extracted code body from the markdown-formatted text returned by LLMs and cleaned non-function parts, such as execution examples. 
For the LeetCode subset, we verified functional correctness by executing each generated solution against the platform-provided test cases, achieving pass@1 of 95.6\% (\texttt{GPT-5}), 94.5\% (\texttt{Claude-4.6}), 95.6\% (\texttt{Deepseek-v4}), 95.5\% (\texttt{Gemini 3}), and 95.0\% (\texttt{Qwen 3.5}). Solutions that failed any test case were excluded. For the WoC subset, no executable test suites are available, as functions extracted from production repositories lack self-contained test. However, the consistently high pass rates on LeetCode, along with those observed on MBPP and HumanEval reported below, indicate that LLMs maintain reliable correctness on tasks of comparable complexity, providing no evidence of systematic degradation on WoC-sourced prompts.
We then applied the readability model to LLM-generated code and the corresponding human-written baseline. After that, we compared the readability between code authored by LLMs and humans, as well as between code generated by different LLMs. 

Regarding the prompt set B (RQ3), to conserve computing resources, we selected \texttt{claude-4.6} for further experimentation, as it demonstrated the best readability performance among functionally correct solutions (Section \ref{sec:rq1}). We then inputted the 5,248 prompts to obtain the returned markdown-formatted text ($\textit{temperature} = 0$ and $\textit{max\_tokens} = 4,096$), and used similar methods to extract Python code and remove extraneous information. 
We executed the generated code against test cases provided, observing pass@1 of 95.1\% (MBPP) and 96.3\% (HumanEval), and retained those that passed all tests ($N = 5,024$). 
We then evaluated the readability of the generated code using our readability model.

Note that readability constitutes an independent dimension of code quality and remains observable and practically relevant during code review and debugging regardless of functional correctness. Moreover, we compared the readability scores between solutions that passed and failed the test cases using a Mann-Whitney U test~\cite{nachar2008mann} and found no significant difference with $p = 0.27$. Nevertheless, we still restricted our analysis to verified-correct code where feasible, ensuring that our findings more reflect code that practitioners would adopt in production.

\subsection{Thematic Analysis}

To address RQ2, we conducted a comparative thematic analysis~\cite{terry2017thematic, williams2019art} of LLM-generated code and human-written code. We adopted a best-case sampling strategy because it provides a forward-looking perspective on readability issues that may become more relevant in practice, as LLMs continue to evolve and users have access to multiple generated alternatives. This design allows us to identify which readability issues persist even in the most readable outputs available in our experimental setting. Specifically, for each programming task in the prompt set, we selected the output with the highest readability score among evaluated LLMs as the representative LLM-generated snippet for that task. We then paired each representative LLM-generated snippet with the human-written counterpart and randomly sampled 500 code pairs\footnote{The sample size of 500 per source was determined to balance analytical depth with the practical feasibility of manual annotation.} from the WoC-sourced and LeetCode-sourced datasets, respectively. Based on readability scores within each pair, samples were divided into cases where LLM-generated code scored higher than its human-written counterpart (405 WoC, 328 LeetCode) and cases where human-written code scored higher (95 WoC, 172 LeetCode). Two authors then independently examined and annotated the sampled pairs.

Our coding procedure followed a hybrid deductive-inductive approach. For each pair, the annotators first read both LLM-generated and human-written code in full and identified which of the four readability dimensions (detailed in Section \ref{sec:readability}) contributed to the lower readability of the less readable snippet: Textual Features (TF)~\cite{scalabrino2018comprehensive}, Posnett's Features (PF)~\cite{posnett2011simpler}, Buse and Weimer's Features (BWF)~\cite{buse2008metric}, and Dorn's Features (DF)~\cite{dorn2012general}. Note that, since a single snippet could exhibit readability issue on multiple dimensions, dimension-level coding was treated as a multi-label annotation task, allowing annotators to assign more than one readability dimension when appropriate. Subsequently, We constructed an initial deductive coding scheme by synthesizing nine common readability issue patterns from previous studies on code readability~\cite{piantadosi2020does, scalabrino2018comprehensive, johnson2019empirical, palomba2014they, oliveira2024understanding}, as summarized in Table \ref{tab:readability_issue}. The annotators applied these predefined patterns to the less readable snippets in each pair. Similarly, this process was also treated as a multi-label annotation task.

When an observed readability issue could not be adequately captured by any of the nine predefined patterns, the annotators recorded it through open coding~\cite{corbin2014basics}. Each annotator independently assigned a descriptive label to the observed issue, and then two annotators compared their open codes. Labels referring to the same phenomenon were merged under a unified name iteratively through discussion, while labels proposed by only one annotator were reviewed jointly and either retained or discarded by consensus. The retained open codes were further synthesized into additional readability issue patterns distinct from the predefined ones.

To assess inter-rater reliability, we calculated Cohen's Kappa values. Since our coding process allowed multiple labels per snippet, we evaluated agreement separately for each label by converting it into a binary coding decision. For readability-dimension coding, Cohen's Kappa values across the four dimensions ranged from 0.81 to 0.89, with a mean of 0.84, while for issue-pattern coding, Cohen's Kappa values across the predefined issue patterns ranged from 0.73 to 0.85, with a mean of 0.79, indicating substantial inter-rater agreement. All discrepancies were discussed and resolved by consensus to produce the final results.

\begin{table}[t] 
    \centering
    \caption{Dimensions of Prompt Crafting} 
    \label{tab:prompt_dimension}
    \vspace{-.5em}

    \begin{tabularx}{\columnwidth}{>{\footnotesize}l>{\footnotesize}X} 
        \toprule 
        \textbf{Dimension} & \textbf{Explanation} \\
        \midrule 
        \textit{Minimum task description}  & The essential instruction or objective of the prompt, stated in a brief and self-contained form. \\
        \textit{Style description} & Instructions specifying the desired code presentation style, such as formatting, naming, and commenting. \\
        \textit{Function signature} & An instruction specifying the expected function interface, typically including the function name, parameters, and type hints. \\
        \textit{IO Contract} & A precise description of the expected input and output structures, formats, and contents for the generated code.\\
        \textit{Few-shot instruction} & Providing the model with examples of the input/output pairs for the specific task. \\
        \textit{Task category} & Explicitly naming the general type of problem or task that the model is being asked to perform. \\
        \textit{Persona setting} & Assigning the model a specific identity, role, or expertise to adopt during generation. \\
        \textit{Constraints} & Restrictive guidelines specifying properties that the generated code should avoid or minimize. \\
        \bottomrule 
        \vspace{-1em}
    \end{tabularx}
    \vspace{-2em}
\end{table}

\subsection{Statistical Analysis}~\label{sec:method-regression}

To address RQ3, we conducted a statistical analysis of the state vectors of 5,024 prompts in set B and the corresponding readability scores of generated code. 
First, we performed a random forest regression~\cite{breiman2001random} using all prompt dimensions as predictors, with $n\_estimators=100$, $max\_depth=15$, and $min\_samples\_split=5$. Random forest regression is appropriate for our analysis because it can model non-linear associations between prompt dimensions and readability scores, while accounting for interaction effects among dimensions. Using the model-based feature importance estimates~\cite{iranzad2025review}, we identified dimensions with stronger predictive relevance to generated code readability.

We then evaluated the statistical significance of each dimension’s association with code readability through methodological triangulation, combining t-test~\cite{mishra2019application} and permutation importance test~\cite{altmann2010permutation}. The t-tests were used to directly compare the readability scores of code generated under two prompt settings (i.e., baseline prompts vs. prompts with the tested dimension augmented) and to determine whether their score differences were statistically significant (when $p < 0.01$). Permutation importance tests assessed whether the tested dimension contributed to model-based prediction performance. Specifically, we used the random forest model as prediction kernel and randomly shuffled the binary indicator of the tested dimension over $10,000$ iterations. A dimension was considered relevant if preserving its original indicator led to significantly better predictive performance than the shuffled cases (when $p < 0.01$). By synthesizing the evidence, we identified whether each dimension significantly correlates with generated code readability. To further ensure reliability, we conducted an ablation analysis comparing full-dimension prompts with prompts in which one dimension was removed, thereby examining whether the absence of each dimension produced convergent evidence in the opposite direction.

\begin{table}[t] 
    \centering
    \caption{Common Readability Issue Patterns} 
    \label{tab:readability_issue}
    \vspace{-.5em}

    \begin{tabularx}{\columnwidth}{>{\footnotesize}l>{\footnotesize}X} 
        \toprule 
        \textbf{Issue Pattern} & \textbf{Explanation} \\
        \midrule 
        \textit{Deficient Comment (DC)}  & Complex or non-obvious code lacks necessary explanations of its purpose, behavior, or important implementation logic.\\
        \textit{Redundant Comment (RC)} & Comments merely duplicate information that is already clear from the code itself, offering no additional value.\\
        \textit{Inconsistent Style (IS)} & The code fails to maintain a consistent style or follow common conventions in indentation, formatting, or idiomatic usage.\\
        \textit{Excessive Complexity (EC)} & The logic or control flow is overly convoluted, deeply nested, or implemented via unnecessarily complicated mechanisms.\\
        \textit{Poor Structure (PS)} & Code snippets are poorly organized, with unclear responsibility separation or weak grouping of related operations.\\
        \textit{Poor Naming (PN)} &  Names for variables, functions, classes, or other identifiers are vague, misleading, or inconsistent with their purpose or scope.\\
        \textit{Magic Values (MV)} &  The code directly uses raw constant values (e.g., numbers or strings) without explanatory context.\\
        \textit{Code Duplication (CD)} &  Highly similar code fragments appear multiple times within the snippet or program, increasing maintenance cost error risk.\\
        \textit{High Coupling (HC)} &  Components, functions, or modules are excessively dependent on one another, so that a change in one component requires cascading changes in others.\\
        \bottomrule 
        \vspace{-1em}
    \end{tabularx}
    \vspace{-1.2em}
\end{table}

\section{Readability Model}\label{sec:readability}

To evaluate code readability, we constructed a readability assessment model from a broad pool of established readability indicators, followed by feature selection and model construction to derive the final assessment model. Specifically, we first organized candidate features from four commonly studied families of readability-related indicators: TF~\cite{scalabrino2018comprehensive}, PF~\cite{posnett2011simpler}, BWF~\cite{buse2008metric}, and DF~\cite{dorn2012general}.  
This selection is based on extensive code readability literature~\cite{lawrie2006s,scalabrino2018comprehensive,posnett2011simpler,buse2008metric,dorn2012general,scalabrino2016improving}, representing widely adopted indicators of code readability across different programming contexts. 

From this candidate feature space, we developed our readability model through feature selection, model training, and empirical validation, retaining indicators that were empirically informative for readability estimation. The resulting model integrates established readability indicators into a calibrated measurement tool for assessing code readability.

In this section, we focus on the rationale for the candidate feature families and summarize key implementation process and evaluation results. The complete technical details and mathematical definitions are provided in the supplementary appendix.

\subsection{Metric}\label{subsec:Metr}

\subsubsection{TF: Semantic and Lexical Perspective}
To capture semantic and lexical aspect of code readability, we incorporated TF features~\cite{scalabrino2018comprehensive, scalabrino2016improving}. 
This feature family is motivated by the observation that identifiers and comments are primary carriers of developer intent~\cite{lawrie2006syntactic,lawrie2007effective}. 
Accordingly, TF characterizes whether the wording used in code is natural, semantically precise, and internally consistent with the documented behavior of the snippet. 
Representative signals in this family describe naming quality, comment-code consistency, lexical ambiguity, and conceptual cohesion. 
Collectively, these features reflect how easily a reader can relate program text to the underlying problem domain and recover the intended meaning of the code implementation.

\subsubsection{BWF: Structural Formatting Perspective}
We employed BWF features~\cite{buse2008metric} to capture the formatting and layout characteristics. 
This family focuses on low-level structural regularities that can be perceived directly from the arrangement of code on the screen without deep semantic analysis. 
In our setting, BWF summarizes three complementary aspects: the horizontal density of individual lines, the syntactic distribution within lines, and the vertical segmentation induced by blank lines and comments. 
Collectively, these features reflect whether a snippet is visually crowded, unevenly distributed, or clearly partitioned into manageable units, which affect scanning effort and local comprehension~\cite{buse2008metric,woodfield1981effect}.

\subsubsection{PF: Information-Theoretic Perspective}
To assess readability from an information-theoretic perspective, we incorporated PF features based on entropy~\cite{posnett2011simpler}. Code becomes more difficult to process when its information content is high or its token distribution is irregular. This family therefore captures the compactness, diversity, and unpredictability of the textual signal, measuring the overall information volume of a snippet and the degree of lexical disorder between lines. From a cognitive perspective, these measures approximate the amount of information that a reader must organize and retain when reading code~\cite{shannon1948mathematical,halstead1977elements}.

\subsubsection{DF: Visual and Geometric Perspective}
To account for visual and geometric properties, we adopted DF features~\cite{dorn2012general}, which model code as a visual object. This feature family characterizes the shape, alignment, and visual distribution of code on the page, reflecting whether a snippet exhibits a regular spatial layout and consistent alignment of syntactic elements. Such properties influence how efficiently readers can detect visual patterns before fully parsing the text, as spatial regularity and alignment facilitate code navigation~\cite{shamy2023identifying,de2024assessing,busjahn2015eye,crosby2002we}.

Together, these four feature families offer complementary coverage of code readability, reducing the risk of relying on any single class of indicators.

\subsection{Implementation and Evaluation}

To operationalize the metric families described above, we implemented a unified feature extraction pipeline that transforms raw code snippets into snippet-level readability features~\cite{moonen2001generating}. We then instantiated the four feature families by implementing the corresponding indicators according to their definitions. 
All metrics are represented as scalar features at the snippet level and concatenated in a fixed order to form the raw feature vector $\mathbf{x} = [\text{TF}, \text{BWF}, \text{PF}, \text{DF}]$. 
Specifically, TF, BWF, PF, and DF yield 16, 26, 4, and 15 features, respectively, resulting in a 61-dimensional candidate feature space\footnote{Full implementation details are recorded in the appendix, including language-specific extraction patterns, normalization procedures, and feature aggregation mechanisms.}.

To perform feature selection and evaluate our readability model, we conducted experiments on the Dorn dataset~\cite{dorn2012general}, a widely used benchmark for code readability evaluation consisting of 360 code snippets. Each snippet is associated with a binary readability label derived from human expert annotations.

We adopted a 10-fold stratified cross-validation protocol
~\cite{kohavi1995study}. Within each fold, feature standardization was fitted on the training split and then applied to the test split to avoid data leakage. Given the size of the 61-dimensional candidate feature space, we applied Sequential Forward Selection (SFS)~\cite{guyon2003introduction,whitney1971direct} to identify a discriminative subset of features. We utilized a Logistic Regression classifier with L2 regularization as the base estimator for SFS.
SFS was performed only within the training split of each fold, starting from an empty set and iteratively adding features that improved validation performance. Table~\ref{tab:eval_results} reports the optimal number of selected features for each setting.

We report Accuracy and Area Under the ROC Curve (AUC)~\cite{fawcett2006introduction}, which capture classification accuracy and ranking quality, respectively. The evaluation covers the four individual feature families, their concatenation (All-features) and the comprehensive model developed by Linear Forward Selection (LFS) in previous research~\cite{scalabrino2018comprehensive}. As shown in Table~\ref{tab:eval_results}, combining the four feature families provides better predictive power for readability over a single feature family. Moreover, feature selection improves the integrated representation, as All-features+SFS ($n_\textit{features}=25$) achieved the best overall result with 77.5\% accuracy and 83.8\% AUC. The SFS-based settings also performed better than All-features+LFS (72.8\% accuracy and 73.7\% AUC), indicating that SFS is more effective for selecting discriminative readability indicators.

These results indicate that the four metric families provide complementary information and that feature selection is an important step for transforming the raw integrated metrics into a more effective readability assessment model, leading to consistent improvements over individual feature families and previous standalone configurations.

\begin{table}[t]
\centering
\caption{Cross-validated performance of our readability model}
\label{tab:eval_results}
\small
\setlength{\tabcolsep}{4pt}
\begin{tabularx}{\columnwidth}{lccc}
\toprule
\textbf{Feature Family} & \textbf{Features} & \textbf{Accuracy} & \textbf{AUC} \\
\midrule
TF & 16 & 64.2\% & 64.4\%\\
TF+SFS &3 & 64.4\%&68.5\%\\
BWF  & 26 & 72.1\% & 78.7\% \\
BWF+SFS & 10& 72.9\%&80.5\%\\
PF & 4 & 66.7\% & 71.2\%\\
PF+SFS&2&63.3\%&69.3\%\\
DF & 15 & 66.1\% & 75.8\%\\
DF+SFS&10&67.8\%&76.6\%\\
All-features&61&71.7\%&80.5\%\\
All-features+SFS {\footnotesize(default)} & 21 & 74.4\% & 83.4\% \\
All-features+SFS {\footnotesize($n_{\textit{features}}=25$)} & 25 & \textbf{77.5\%} & \textbf{83.8\%} \\
All-features+LFS &14&72.8\%&73.7\%\\

\bottomrule
\end{tabularx}
\vspace{-1em}
\end{table}

\section{RQ1: Readability Comparison}\label{sec:rq1}

\begin{table}[t] 
    \centering
    \caption{Readability Scores of LLM-generated and Human-written Code} 
    \label{tab:rq1_res}
    \resizebox{\columnwidth}{!}{
    \begin{threeparttable}

    \begin{tabular}{ccccc} 
        \toprule 
        \textbf{Model} & \textbf{Data Source} & \textbf{Avg (LLM)} & \textbf{Avg (Human)} & \textbf{Win Rate (\%)}\\ 
        \midrule 
        \multirow{2}{*}{Claude-4.6} & WoC & 0.11 & -0.15 & 55.00 \\
                           & LeetCode & 3.47 & 3.14 & 57.75 \\
        \midrule
        \multirow{2}{*}{GPT-5} & WoC & 0.17 & -0.15 & 55.70 \\
                           & LeetCode & 2.93 & 3.17 & 41.76 \\
        \midrule
        \multirow{2}{*}{Gemini 3} & WoC & 0.53 & -0.15 & 64.70 \\
                           & LeetCode & 3.27 & 3.14 & 50.29 \\
        \midrule
        \multirow{2}{*}{DeepSeek-v4} & WoC & 0.31 & -0.15 & 60.60 \\
                           & LeetCode & 3.21 & 3.17 & 49.50 \\
        \midrule
        \multirow{2}{*}{Qwen 3.5} & WoC & -0.05 & -0.15 & 51.30 \\
                           & LeetCode & 3.29 & 3.14 & 55.10 \\
        \midrule
        \multirow{2}{*}{All} & WoC & 0.22 & -0.15 & 57.46 \\
                           & LeetCode & 3.25 & 3.15 & 51.15 \\
        \bottomrule 
        \vspace{-1em}
    \end{tabular}
    \begin{tablenotes}
            \footnotesize 
            \item[*] \textbf{Avg} refers to the average readability score of corresponding code snippets evaluated by the readability model; \textbf{Win Rate} refers to the proportion of samples where LLM-generated code strictly outperforms the human-written counterpart in readability.
            \item[*] \textbf{Avg(Human)} scores on LeetCode vary slightly across models because instances where LLM-generated counterpart fails the test are excluded from the calculation.
        \end{tablenotes}
    
    \end{threeparttable}
    }
    \vspace{-.5em}
\end{table}

Table~\ref{tab:rq1_res} shows that LLM-generated code is generally comparable to, and in many settings slightly more readable than, human-written code. In WoC, all models achieve higher average readability scores than the human-written counterparts, with win rates ranging from 51.30\% to 64.70\%. While in LeetCode, the advantage is weaker: several models show only marginal advantages, and GPT-5 falls below the human baseline in both the average score and win rate. Note that although WoC samples exhibit lower absolute scores due to more complex contexts inherent in real-world software tasks, our comparison remains within-task, with each LLM output compared against the corresponding human-written code.

This difference suggests that the relative readability of LLM-generated code depends on the task setting. LeetCode problems are usually self-contained and well-specified~\cite{xia2025leetcodedataset}, and human programmers often have accumulated stable solution patterns for such tasks~\cite{meerbaum2011habits}, leading to relatively concise and readable implementations and leaving limited room for LLM-generated code to improve. By contrast, WoC samples come from more complex real-world software contexts~\cite{ma2019world}, where implementation details, dependencies, and functionality-first practices~\cite{meerbaum2011habits} may make human-written code less uniformly readable. LLMs may benefit from style regularities learned during pretraining~\cite{chen2021evaluating}, producing code that appears more standardized and readable under our assessment model. However, since WoC snippets are not accompanied by executable test suites, this readability advantage should not be interpreted as evidence of functional correctness.

\begin{figure}[htbp]
    \centering
    \subfloat[Boxplot of the readability scores\label{subfig:llm_human_boxplot}]{
        \includegraphics[width=0.35\textwidth]{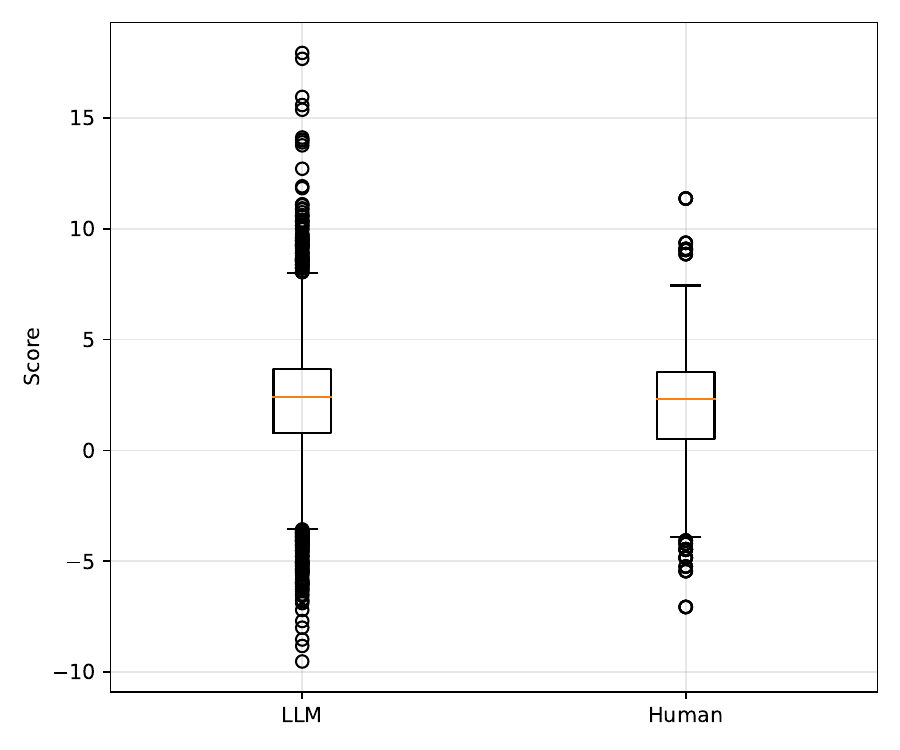}
    }
    \hfill
    \subfloat[Histogram of the readability scores\label{subfig:llm_human_histogram}]{
        \includegraphics[width=0.35\textwidth]{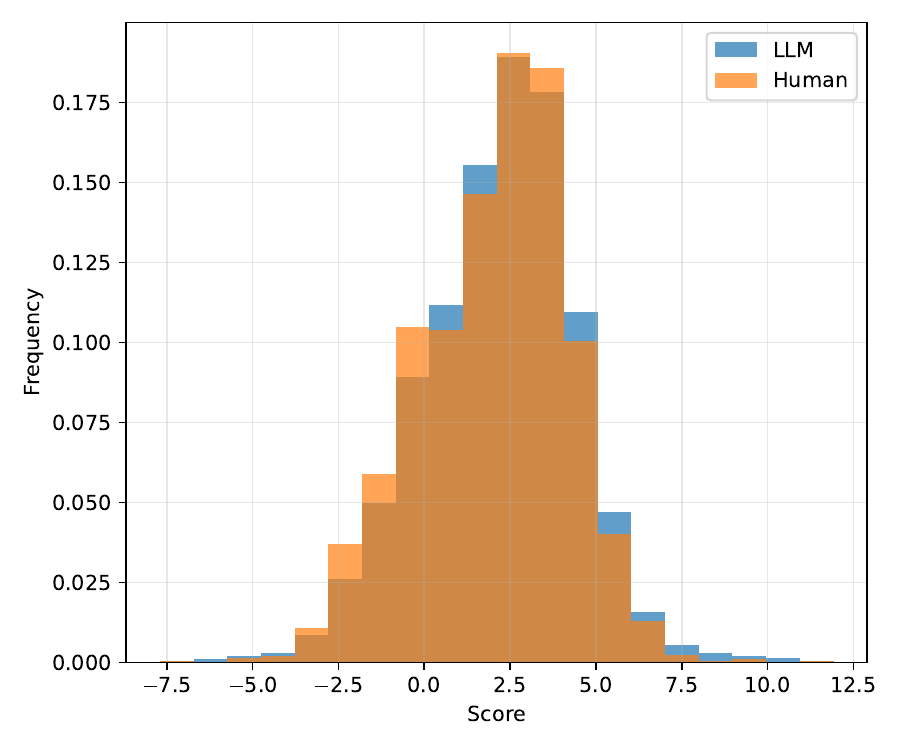}
    }
    \caption{Comparison of the readability scores of LLM-generated code and Human-written code}
    \label{fig:llm_human}
\end{figure}

Figure~\ref{fig:llm_human} further illustrates the readability score distributions, revealing that while LLMs achieve a marginally higher median, human-written code demonstrates better consistency, as evidenced by a tighter interquartile range and fewer outliers (Figure~\ref{subfig:llm_human_boxplot}). This suggests that LLM-generated code shows more variation across tasks, which may be attributed to LLMs' inherent next-token prediction mechanism~\cite{vaswani2017attention}, in contrast to programmers' established programming habits~\cite{meerbaum2011habits}. Figure~\ref{subfig:llm_human_histogram} (15 extreme outliers are excluded for better visualization) intuitively shows substantial distributional overlap between LLM-generated and human-written code.

To establish a more rigorous comparison, we applied the Wilcoxon signed-rank test~\cite{wilcoxon1945individual} to all matched LLM-human pairs, thereby controlling for task-specific variance. The result reveals a statistically significant difference between two groups ($p < 0.001$), but the effect size is small ($r = 0.104$), especially given the large number of paired samples ($N=13{,}263$). Thus, the statistical evidence supports a small but detectable difference rather than a substantial readability gap.

\textbf{Taken together, these results suggest that LLM-generated code and human-written code are generally comparable in readability.} The slight advantage observed for LLM-generated code is more evident in WoC than in LeetCode, indicating that readability differences vary across benchmark contexts with different task settings.

\begin{figure}[htbp]
    \centering
    \subfloat[WoC-sourced prompt results\label{subfig:diff_llm_woc}]{
        \includegraphics[width=0.45\textwidth]{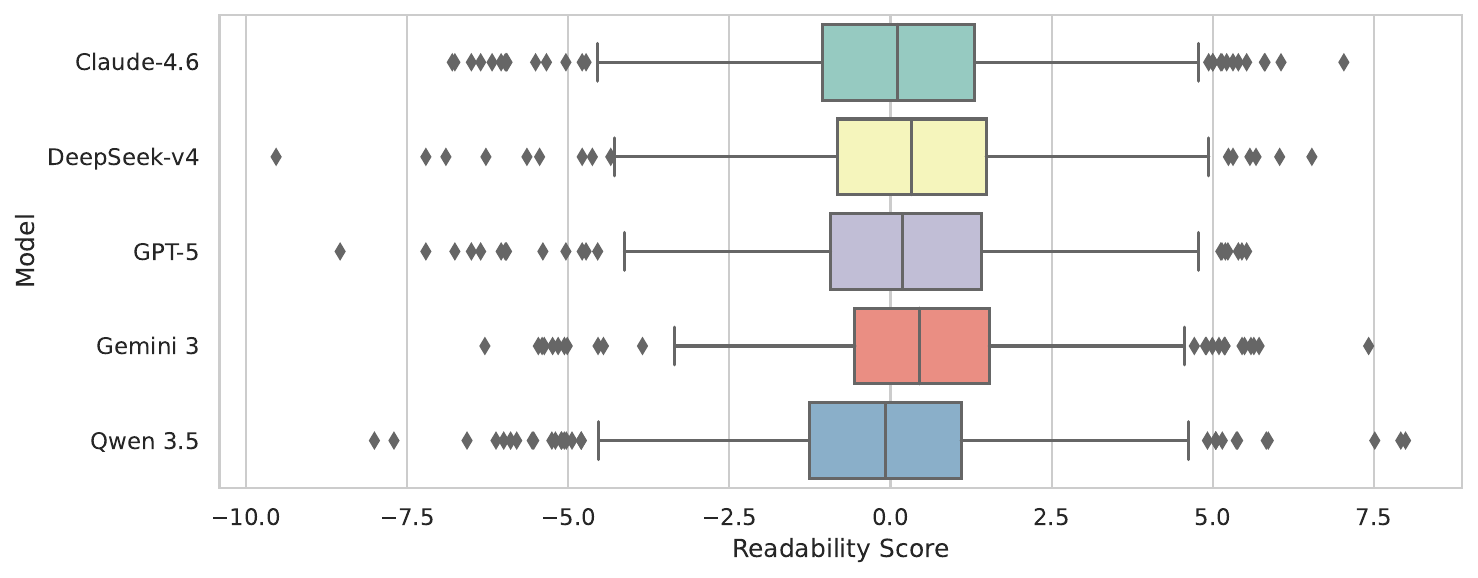}
    }
    \hfill
    \subfloat[LeetCode-sourced prompt results\label{subfig:diff_llm_leetcode}]{
        \includegraphics[width=0.45\textwidth]{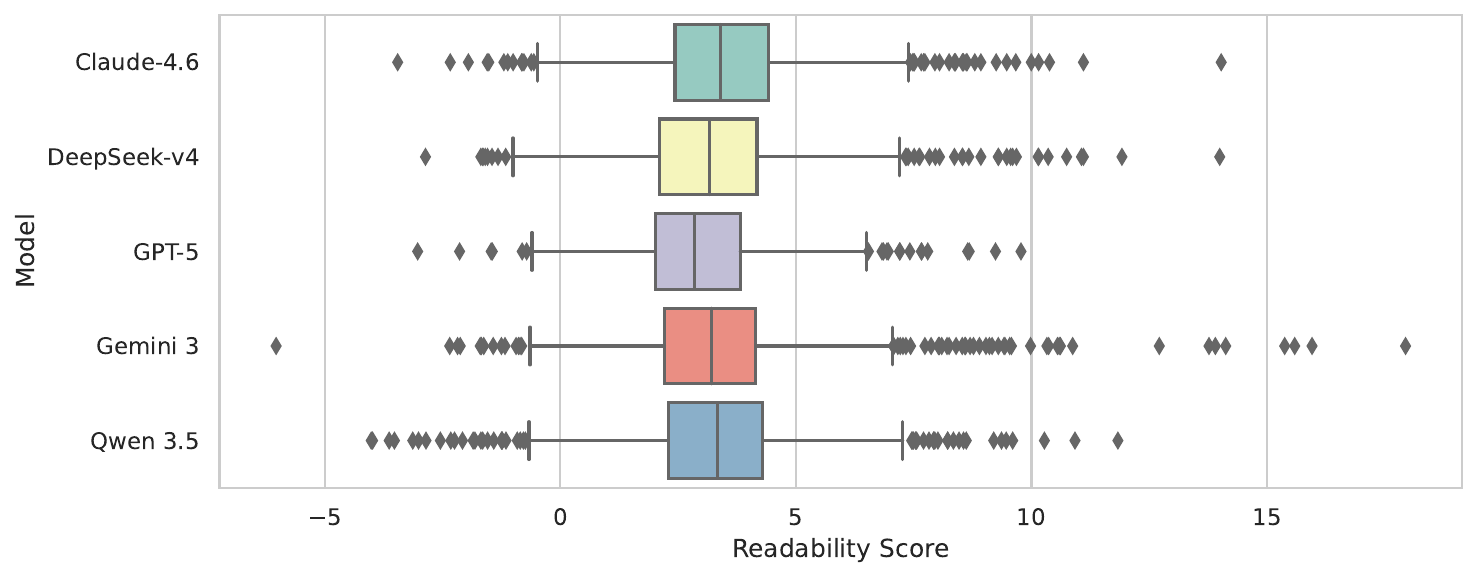}
    }
    \caption{Readability score distribution of code generated by different LLMs}
    \label{fig:diff_llm}
\end{figure}

Beyond the overall LLM-human comparison, the 2,735 prompt-code pairs also provide a benchmark enabling a cross-model evaluation of code readability. As shown in Figure~\ref{fig:diff_llm}, \texttt{Claude-4.6} achieves the strongest readability performance on functionally verified LeetCode code and maintains a stable overall readability profile across the evaluated settings. We therefore use \texttt{Claude-4.6} in the experimental setting (RQ3, Section \ref{subsec:prompt}) as a strong representative model.

\section{RQ2: Readability Issue Patterns}\label{sec:rq2}

Table~\ref{tab:issue_dimension} shows that less readable human-written and LLM-generated code exhibit different dimension-level patterns. Human-written snippets mainly suffer from BWF and TF issues, indicating problems in formatting, layout, comments, and textual expression, while PF is much less frequent. In contrast, less readable LLM-generated snippets are dominated by TF and PF issues, suggesting problems in semantic expression and unnecessary information load rather than surface formatting.

These distinct patterns reflect the underlying differences in code generation context. Human programmers often prioritize functional correctness and development efficiency over stylistic rigor and considerate textual expression, leading to BWF and TF issues~\cite{rebouccas2023s,lenarduzzi2021systematic}. However, their experience-driven reliance on validated concise logic typically results in lower information entropy and predictable structures, thereby minimizing PF issues~\cite{meerbaum2011habits}. In contrast, having been trained on substantial amounts of standardized code~\cite{vaswani2017attention}, LLMs tend to generate style-standard code with less BWF issues, but their lack of experiential best practice often leads to correct but over-engineered solutions, manifesting as PF logic bloat~\cite{licorish2025comparing, zi2025would}. Furthermore, TF emerged as the most frequent issue dimension in LLM-generated code, which could be attributed to the lack of emphasis on semantic appropriateness in training data and model’s tendency to hallucinate text that is irrelevant to the actual code logic~\cite{ji2023survey,kang2024identifying}. These patterns indicate that LLM and human code can have similar readability scores for different underlying reasons.

The issue categories in Table~\ref{tab:issue_category} make these dimension-level patterns more concrete, showing which observable coding problems give rise to the dimensional concerns in each group. In human-written code, the most frequent issue patterns are Deficient Comments (DC) and Inconsistent Style (IS), which align with the high TF and BWF counts in Table~\ref{tab:issue_dimension}, while Redundant Comments (RC) and High Coupling (HC) are rare. This pattern is consistent with the tendency of programmers to prioritize functional correctness and to speed over non-essential tasks like documentation and styling~\cite{lenarduzzi2021systematic}. Regarding comments as secondary to execution, developers often omit necessary explanations or simplify them too much, which leads to high DC but very few RC issues~\cite{rebouccas2023s}.

The most frequent issue in LLM-generated code is Excessive Complexity (EC), followed by Redundant Comments (RC), which align with the high TF and PF counts in less readable LLM-generated snippets. Unlike humans, LLMs do not inherently aim for a simple solution. Instead, they may produce code with extra intermediate steps, over-verbose control flow, or unnecessary explanations, making the code harder to follow even when its formatting is clean and function is correct~\cite{licorish2025comparing,zi2025would, kang2024identifying}. The low incidence of Code Duplication (CD) suggests that LLMs are good at modularizing code and reusing logic without repetition. High Coupling (HC) does not appear in either group, likely due to the function-level granularity of our snippets. We further discuss it in Section~\ref{sec:discussion}.

\begin{table}[t] 
    \centering
    \caption{Distribution of Issues on Readability Dimensions} 
    \label{tab:issue_dimension}
    \vspace{-.5em}

    \begin{tabular}{lcccc} 
    \toprule 
        \textbf{Category} & \textbf{TF} & \textbf{BWF} & \textbf{PF} & \textbf{DF} \\
        \midrule 
        Human-bad & 72&\textbf{76}&\uline{21}&45\\
        LLM-bad &\textbf{60}&\uline{26}&53&30 \\
        Total & \textbf{132}&102&\uline{74}&75 \\
        \bottomrule
    \end{tabular}
    \par\vspace{.5em}
    \parbox{0.95\columnwidth}{
            \scriptsize \textsuperscript{*}\hspace{.2em}TF (Semantic/Lexical) evaluates intent and naming; BWF (Structural/Formatting) assesses layout and density; PF (Information-Theoretic) measures logical complexity and entropy; DF (Visual/Geometric) captures spatial alignment.
        }
\end{table}

\begin{table}[t] 
    \centering
    \caption{Categories of Readability Issue Patterns} 
    \label{tab:issue_category}
    \vspace{-.5em}
    \resizebox{\columnwidth}{!}{
    \begin{tabular}{lccccccccc} 
    \toprule 
        \textbf{Category} & \textbf{DC} & \textbf{RC} & \textbf{IS} & \textbf{EC} & \textbf{PS} & \textbf{PN} & \textbf{MV} & \textbf{CD} & \textbf{HC}\\
        \midrule 
        Human-bad & \textbf{60}&\uline{3}&\textbf{58}&41&24&22&7&6&0\\
        LLM-bad & 10&\textbf{24}&15&\textbf{48}&16&21&8&\uline{2}&0 \\
        Total & 70&27&\textbf{73}&\textbf{89}&40&43&15&\uline{8}&0 \\
        \bottomrule

    \end{tabular}
    }
\end{table}

Beyond predefined issue categories, open coding revealed three additional issues patterns in LLM-generated code: Unknown API, Redundant Variables, and Overblanking. Unknown API occurs when LLMs introduce external library calls without sufficient context, no matter whether these APIs are hallucinations or real. Even if the API is valid, its presence can make the snippet harder to understand when the dependency is unfamiliar or its role is not explained~\cite{robillard2011field, wang2024llms}. Redundant Variables refer to cases where multiple variables serve the same functional purpose, creating unnecessary indirection and making the data flow harder to trace. Overblanking involves excessive but meaningless blank lines that lengthen the code without improving structural separation. Although related to formatting, we treat it separately from IS and PS because it reflects a different problem, over-segmentation without a semantic boundary, instead of stylistic inconsistency or flawed logical organization.
Its repeated occurrence in LLM-generated code suggests a surface-level use of spacing that does not always align with the logical structure.

\textbf{In summary, LLM-generated and human-written code exhibit distinct readability failure modes: human-written code more often lacks explanation and consistent presentation, while LLM-generated code more often suffers from redundant explanation and unnecessary complexity.}

\section{RQ3: Role of Prompt Design}\label{sec:rq3}

\begin{figure}[b]
    \centering
    \includegraphics[width=0.95\linewidth]{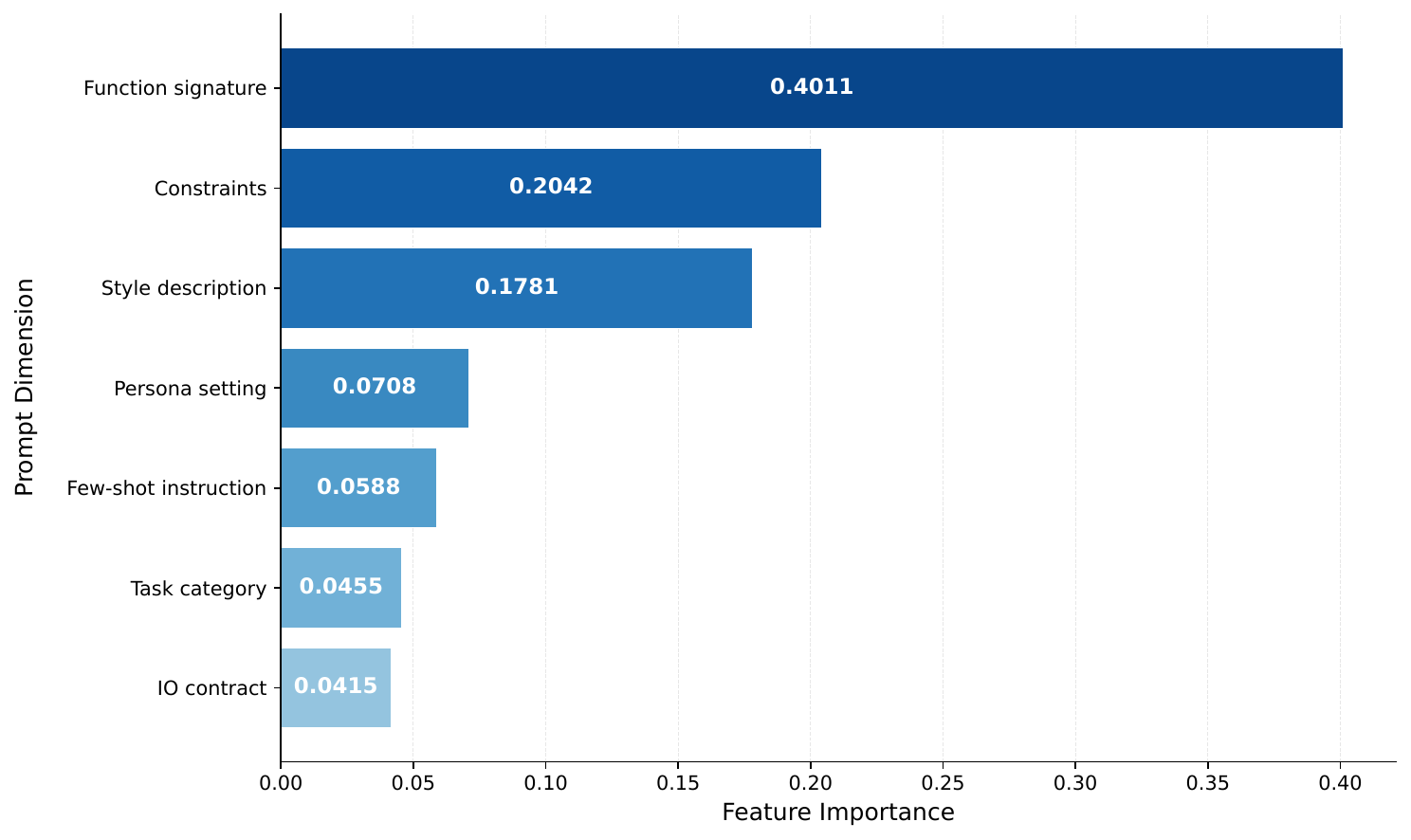}
    \caption{Feature importance of the Random Forest regression result }
    \label{fig:randomforest}
\end{figure}

Figure~\ref{fig:randomforest} reports the feature importance scores of the random forest regression, providing an overall ranking of prompt dimensions by their predictive relevance to generated code readability.
\textit{Function signature} shows the highest importance among all evaluated dimensions, followed by \textit{constraints} and \textit{style description}, while \textit{task category} and \textit{IO contract} contribute the least. This suggests that prompt components closer to the form of the generated code are more associated with readability. \textit{Function signatures} define the interface, naming context, and expected structure of a function-level snippet~\cite{ding2024code}, while \textit{style descriptions} and \textit{constraints} provide explicit guidance, offering \textit{``dos and don'ts''}.
In contrast, \textit{task category} and \textit{IO contract} mainly specify what problem to solve and how inputs and outputs should behave, which are more directly related to functional logic instead of readability. 

The $R^2$ of the random forest model ($<0.3$) further contextualizes this ranking. Prompt dimensions explain only part of the readability variation, while other factors, such as problem complexity, model capability, and generation variability, may also play substantial roles~\cite{alawad2019empirical,licorish2025comparing}. We therefore interpret prompt dimensions as lightweight levers for improving readability, rather than as dominant determinants.

\begin{table}[t]
    \centering
    \caption{P-value of Statistic Analysis of Prompt Dimensions and Code Readability}
    \label{tab:rq3_res}
    \setlength{\tabcolsep}{0pt} 
    \begin{threeparttable}
        \begin{tabular*}{\columnwidth}{@{\extracolsep{\fill}} l cccc @{}}
            \toprule
            \multirow{2}{*}{\textbf{Dimension}} & \multicolumn{2}{c}{\textbf{Incremental}} & \multicolumn{2}{c}{\textbf{Ablation}} \\
            \cmidrule(lr){2-3} \cmidrule(lr){4-5}
            & T-test & Permutation & T-test & Permutation \\
            \midrule
            \textit{Style description}   & \textbf{0.001} & \textbf{0.001} & 0.795 & 0.850 \\ \addlinespace[0.5em]
            \textit{Function signature}  & 0.239 & 0.240 & \textbf{0.003} & \textbf{0.003} \\ \addlinespace[0.5em]
            \textit{IO Contract}         & 0.353 & 0.356 & 0.855 & 0.887 \\ \addlinespace[0.5em]
            \textit{Few-shot}            & 0.257 & 0.291 & 0.643 & 0.678 \\ \addlinespace[0.5em]
            \textit{Task category}       & 0.282 & 0.322 & 0.673 & 0.684 \\ \addlinespace[0.5em]
            \textit{Persona setting}     & 0.287 & 0.282 & 0.281 & 0.287 \\ \addlinespace[0.5em]
            \textit{Constraint}          & \textbf{0.001} & \textbf{0.000} & 0.807 & 0.960 \\
            \bottomrule
        \end{tabular*}
        \begin{tablenotes}
            \footnotesize
            \vspace{.3em}
            \item[*] Bold values indicate statistical significance ($p < 0.01$).
        \end{tablenotes}
    \end{threeparttable}
    \vspace{-1em}
\end{table}

Table~\ref{tab:rq3_res} further shows which prompt dimensions show statistically significant associations with code readability. In the incremental setting, where each dimension is added to a baseline prompt, \textit{style descriptions} and \textit{constraints} show significant positive associations with code readability. This suggests that when starting from a minimal task description, explicitly declaring readability-oriented guidance is useful, as \textit{style descriptions} specify the desired form of readable code while \textit{constraints} restrict undesirable forms. 

The ablation setting provides a complementary perspective. When one dimension is removed from comprehensive prompts, only removing \textit{function signature} leads to a significant change in code readability. This suggests that, in well-designed prompts, \textit{function signature} plays a distinctive role in shaping code readability. Beyond merely adding guidance, signatures provide a structural anchor for generation by specifying the function interface, naming context, and expected organization of the snippet~\cite{ding2024code}. When other prompt dimensions are already present, removing this anchor has a clearer readability cost than removing other dimensions.
 
\textbf{In summary, the analyses show a consistent but bounded role of prompt design, with \textit{function signatures}, \textit{constraints}, and \textit{style descriptions} emerging as the most relevant prompt dimensions associated with code readability.
\textit{Style description} and \textit{constraints} are most useful when the prompt is underspecified, whereas \textit{function signature} is most important when the prompt is otherwise complete. These dimensions offer practical targets for low-cost readability improvements, but the modest explanatory power of the random forest model cautions against treating prompt engineering as a complete solution.} 

\section{Discussion}\label{sec:discussion}

\subsection{Multi-Turn Interactions and Readability}

Our study adopts a single-turn prompting setting, which does not fully capture the iterative, multi-turn interactions increasingly common in practical workflows. However, single-turn readability remains important in such settings, and we discuss the relevance of our findings to multi-turn readability refinement.

\textbf{Readability as a human-perceived intermediate property.} Unlike functional correctness, which can be checked automatically through test execution, readability is tied to human cognitive effort. In a multi-turn interaction, developers typically need to read the generated code at each turn before they can formulate meaningful refinement instructions. Thus, a multi-turn session can be regarded as a sequence of intermediate reading episodes. The comprehension cost incurred at each intermediate turn accumulates throughout the interaction and is not offset by an improved final output. Therefore, single-turn readability characterizes not merely an isolated snapshot but a recurring cost that developers may face at every iteration of a multi-turn workflow.

\begin{figure}[t]
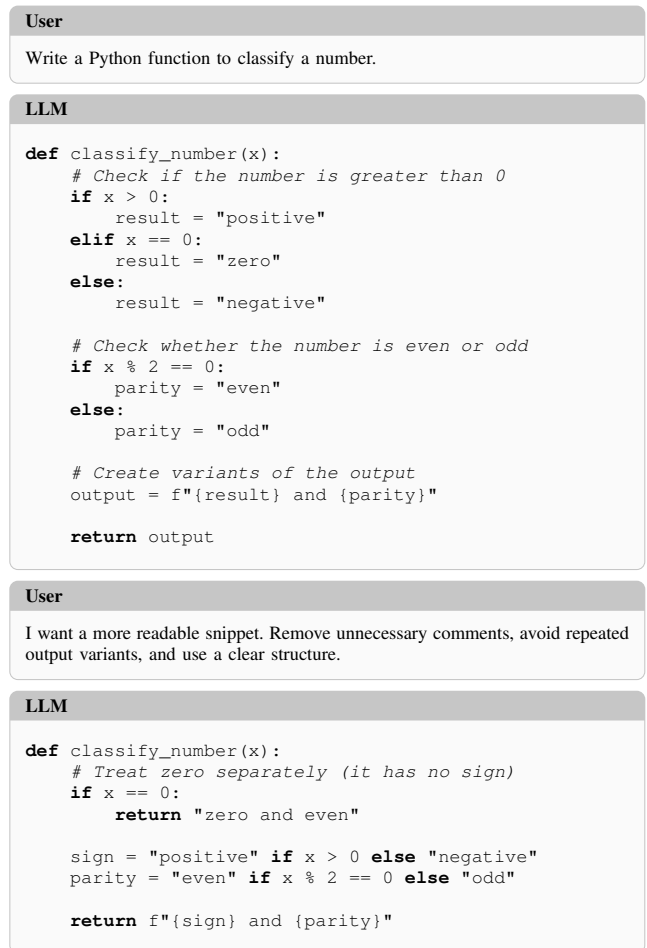

\centering
\begin{minipage}{0.95\linewidth}

\begin{dialoguebox}{User}
\scriptsize
Write a Python function to classify a number.
\end{dialoguebox}

\begin{dialoguebox}{LLM}
\begin{lstlisting}[style=pythonstyle,language=Python]
def classify_number(x):
    # Check if the number is greater than 0 
    if x > 0:
        result = "positive"
    elif x == 0:
        result = "zero"
    else:
        result = "negative"

    # Check whether the number is even or odd
    if x % 2 == 0:
        parity = "even"
    else:
        parity = "odd"

    # Create variants of the output
    output = f"{result} and {parity}"

    return output
\end{lstlisting}
\end{dialoguebox}

\begin{dialoguebox}{User}
\scriptsize
I want a more readable snippet. Remove unnecessary comments, avoid repeated output variants, and use a clear structure.
\end{dialoguebox}

\begin{dialoguebox}{LLM}
\begin{lstlisting}[style=pythonstyle,language=Python]
def classify_number(x):
    # Treat zero separately (it has no sign)
    if x == 0:
        return "zero and even"

    sign = "positive" if x > 0 else "negative"
    parity = "even" if x % 2 == 0 else "odd"

    return f"{sign} and {parity}"
\end{lstlisting}
\end{dialoguebox}

\end{minipage}
\caption{An example of multi-turn refinement process in code generation}
\label{fig:example-interaction}
\end{figure}

\textbf{Illustrative example: multi-turn refinement of readability.} To illustrate how multi-turn interaction may affect readability, we present an example in which we issue explicit follow-up instructions requesting readability improvements. 
Figure~\ref{fig:example-interaction} suggests that targeted follow-up instructions can improve readability to some extent. However, the second-turn instruction is not independent of the first-turn code, which is formulated after inspecting the initial output and identifying what needs to be improved. This reinforces our preceding argument that such instructions are usually informed by the developer's review of the preceding output, with reading cost already incurred.

Thus, single-turn readability remains a key prerequisite for effective multi-turn interaction. This also suggests a potential direction for future work to combine automated detection of readability issues with the generation of refinement instructions, so that the comprehension burden currently placed on developers can be reduced.

\subsection{Implications}

Our findings show that readability deserves attention in LLM-generated code, as it affects how generated code is evaluated, reviewed, and refined.

\textbf{Extending model evaluation beyond correctness.} 
Our study shows that LLM-generated code indeed present readability issues. 
Current benchmarks for code-generating LLMs mainly focus on functional correctness and task completion rate through unit tests~\cite{wang2023far, lu2021codexglue}. However, passing tests does not guarantee that generated code is easy to read, review, or maintain. Models that generate functionally correct but cryptic code can still impose substantial cognitive burdens on developers during review and maintenance~\cite{jaspan2023defining}. Future benchmarks should therefore evaluate generated code not only by whether it passes tests, but also by whether it supports human comprehension. For model developers, incorporating readability into evaluation may also support training and alignment procedures that favor maintainable code beyond merely executable code.

\textbf{Managing readability debt in code review.} Existing review and static analysis tools were largely designed around human-written code~\cite{panichella2015would}, but this work reveals that LLM-generated code presents different readability risks. Human readability issues often involve neglect, such as \textit{Deficient Comments} and \textit{Inconsistent Style}, whereas LLM-generated code more often involves overproduction, such as \textit{Redundant Comments} and \textit{Excessive Complexity}. These issues can accumulate into what we call readability debt: code that looks acceptable at generation time but increases future comprehension and maintenance cost. Code review tools should therefore include checks tailored to LLM-typical problems, detecting logical bloat, unnecessary variable redundancies, unexplained API usage, etc. Integrating these checks into CI/CD workflows can help prevent readability debt from entering the codebase while preserving the productivity benefits of AI-assisted coding.

\textbf{Prompt design as a lightweight intervention.} Prompt design offers a low-cost way to improve readability. Our results suggest that \textit{function signature}, \textit{style description}, and \textit{constraints} are the most actionable dimensions. \textit{Style description} and \textit{constraints} help underspecified prompts, while \textit{function signature} matters when prompts are otherwise complete. Practitioners can use these dimensions as a starting point for readability-oriented prompting. However, since prompt design explains only part of readability variation, stronger improvements may require alternative approaches, such as post-processing~\cite{petersen2021post}, readability-oriented fine-tuning~\cite{wei2021finetuned}, or hybrid human-AI collaborative coding workflows~\cite{hassany2024authoring}.

\textbf{Tool support for readability refinement.} In multi-turn AI-assisted development, developers often need to read generated code before they can write useful refinement instructions~\cite{mozannar2024reading}. This makes readability an interaction cost, not only a final-output property. We therefore advocate future tools that detect readability issues in generated snippets and produce in-context refinement suggestions based on the detected problems. For example, if a snippet contains excessive logic or unexplained API usage, the tool could suggest targeted follow-up prompts to simplify the control flow or clarify the role of API. Such support could reduce the diagnostic burden on developers while keeping them in control of final code quality.

\subsection{Threats to Validity}

\textbf{Internal Validity.}
A key threat concerns the assessment of code readability. Although our readability model integrates four complementary feature families, applies feature selection, and achieves the best performance in the evaluation, readability remains subjective and multifaceted. The model may not fully capture individual developer preferences or project-specific readability norms. Furthermore, our focus on function-level snippets limits the detection of issues such as \textit{High Coupling}, which typically manifests at the file or project level. So its absence in our findings should be interpreted as a result of our analysis granularity rather than evidence that such issues do not occur in LLM-generated code.
Finally, for WoC and HumanEval, we converted docstrings into natural language task descriptions. Although this transformation followed strict constraints and underwent cross-validation, it may still have subtly affected the LLM outputs. Future work could compare outputs generated from original docstrings and transformed prompts to quantify this effect.

\textbf{External Validity.}
One threat concerns our sampling strategy for analyzing readability issue patterns and the role of prompt design. For issue patterns, we selected the most readable LLM output among the evaluated models for each task, while for prompt design, we focused on \texttt{Claude-4.6}, the best model on functionally verified LeetCode code in our comparison. Thus, the corresponding findings should not be interpreted as describing average LLM-generated code. Rather, they characterize readability issues and prompt design associations under a strong LLM-output setting. This design may overestimate LLM readability, but it is appropriate for studying the residual readability weaknesses of strong models as LLM capabilities continue to evolve. 

Another threat concerns the scope of our experimental setting. Our study focuses on Python and a selected set of LLMs, and the findings may differ for programming languages with different syntactic and stylistic conventions~\cite{stefik2013empirical} or for models specifically tuned for code generation. In addition, although our sample includes diverse sources, independent snippets cannot fully capture production environments, where code generation is shaped by project-specific constraints, broader dependencies, and collaborative workflows. Future work is expected to examine readability in larger, project-level human-AI development scenarios.

\section{Conclusion}\label{sec:conclusion}

This study examined the readability of LLM-generated code, its issue patterns, and the role of prompt design. Our results show that LLM-generated code is generally comparable to human-written code in overall readability, but the two differ in how readability breaks down. Human-written code more often suffers from \textit{Deficient Comments} and \textit{Inconsistent Style}, whereas LLM-generated code more often exhibits \textit{Excessive Complexity}, \textit{Redundant Comments}, and additional LLM-specific issues such as \textit{Unknown API usage}, \textit{Redundant Variables}, and \textit{Overblanking}. 
We also find that prompt design can shape readability, with \textit{function signature}, \textit{style description}, and \textit{constraints} being the most relevant dimensions. However, its overall role remains bounded, suggesting that prompt engineering is useful but incomplete. These findings highlight the need to evaluate LLM-generated code not only by functional correctness, but also by the concrete readability issues that may accumulate into latent technical debt. They also suggest that prompt design can serve as a lightweight starting point for readability improvement, while motivating future automated support for detecting, reviewing, and refining readability problems in AI-assisted development.

\bibliographystyle{IEEEtran}

\end{document}